%Paper: hep-ph/9209201
%From: KONIG@physics.carleton.ca
%Date: Tue, 1 Sep 1992 13:37:26 -0400 (EDT)

\magnification=\magstep1
\overfullrule=0pt\nopagenumbers
\hsize=15.5truecm\vsize=21.5truecm
\line{\hfil OCIP/C-92-2}
\centerline{\bf MINIMAL SUPERSYMMETRIC STANDARD MODEL}
\centerline{\bf EXTENDED BY ONE ADDITIONAL HIGGS SINGLET}
\centerline{\bf AND ITS INFLUENCE ON THE ANOMALOUS MAGNETIC}
\centerline{\bf MOMENT OF THE MUON\footnote*
{{\sevenrm Talk presented at Beyond the Standard Model III, June 22-24, 1992,
Ottawa, Canada.}}}
\vskip.2cm
\centerline{H. K\"ONIG \footnote*{{\sevenrm supported by Deutsche
Forschungsgemeinschaft}}}
\centerline{Ottawa-Carleton Institute for Physics}
\centerline{Department of Physics, Carleton University}
\centerline{Ottawa, Ontario, Canada K1S 5B6}
\vskip.2cm
\centerline{ ABSTRACT}\vskip.2cm\indent
\hfill\vbox{\hsize12.9 truecm
We present the results of the influence of the minimal
supersymmetric standard model extended by an additional
 Higgs singlet N, with vacuum expectation
value $v_N$, on the anomalous magnetic moment of the muon.
 This gives different mass matrices for the charginos
and neutralinos, which are taken into account within the relevant penguin
 diagrams leading to a
contribution $\Delta a_{\mu}$\ to the anomalous magnetic moment of the
muon. We show that a large vacuum expectation value for the Higgs singlet leads
to a suppression of the supersymmetric contribution making it difficult
to see in an experiment in the near future.}\hfill
\hfill\break
\vskip.2cm{\bf I. INTRODUCTION}\hfill\break\vskip.2cm\noindent
The anomalous magnetic moment of the Muon (AMMM)
$a_{\mu}={1 \over 2}( g_{\mu}-2)$\ is one of the most precisely known
quantities.
The experimentally measured deviation $\Delta a_{\mu}$\ of $a_{\mu}$\ is within
a range of
$-2\cdot 10^{-8}\le \Delta a_{\mu}^{\rm exp.}\le +2.6\cdot 10^{-8}\ ^1$.
Because of this highly precise measured value the AMMM has become a favorite
parameter, which can be used to look for influences of models beyond
the standard model.\hfill\break\indent
In the near future there will be an
experiment$^2$  which  will give a value for $\Delta a_{\mu}^{\rm exp}$\ about
a
factor 10-20 better than this one above
and thus it becomes more important to look for all
possible contributions to this value, especially in models beyond the
standard model.\hfill\break\indent
In this talk we present how an additional Higgs singlet N
in the superpotential which gets a vev affects the contribution
$\Delta a_{\mu}$\ to the AMMM within the minimal supersymmetric standard model
(MSSM).
 In the first section we describe the MSSM extended by one additional Higgs
singlet.
 We take into account the experimental
mass bounds for charginos, neutralinos, the scalar muon and the scalar muon
neutrino given by the study of $Z^0$\ decays at LEP and SLC$^3$. In the last
section we present the results of the influence of those particles on
the AMMM.\hfill\break\indent
Detailed calculations of the relevant feynman diagrams have been given
elsewhere$^{4,5}$\ and will not be repeated
here.\hfill\break\vskip.2cm\noindent
 {\bf II. THE MINIMAL SUPERSYMMETRIC STANDARD MODEL EXTENDED BY
ONE ADDITIONAL HIGGS SINGLET}\vskip.2cm
In the MSSM the superpotential$^6$ $g_{S3}$\ is described by trilinear
Yukawa couplings of two Higgs doublets to the scalar quark and lepton
doublets and the scalar partner
of the righthanded  quark and lepton singlets.
 The scalar potential$^{7,8}$\ consists of
$F_i^*=-{{\partial g_{S3}}\over{\partial\Phi^i}}$-terms and $D_a=g_a\Phi^{i*}
T^{aj}_i\Phi_j$-terms, soft SUSY breaking mass terms for the scalar
particles, and trilinear scalar couplings of the order of the W-boson mass
for phenomenological reasons.
To get the correct $SU(2)\otimes U(1)$\ breaking$^9$\ and to
avoid an axion$^{10}$\ it is well known that one has to introduce a further
bilinear term in the superpotential $g_{S2}=+\mu H_1\epsilon H_2$.
Because the fermion masses are given by ${\cal L}=-{1\over 2}{{\partial^2 g_S}
\over{\partial\Phi_i\partial\Phi_j}}\Psi_i\Psi_j+h.c.$\ this term also plays an
important role in the mass matrices of charginos and neutralinos$^7$.
In an $E_6$-based superstring inspired model such a mass term
for the Higgs particles is naturally
absent$^{11}$. To avoid this term and still to get a
phenomenologically possible supersymmetric (SUSY) model with only trilinear
terms, a Higgs singlet N was introduced$^{8,12}$.
\hfill\break\indent In
the following, we extend our superpotential $g_{S3}$\ by the
terms $g_N=h_N H_1\epsilon H_2N-{1\over 3}kN^3$. The $kN^3$\ is
important for avoiding an unacceptable axion according to  a global U(1)
symmetry. Detailed calculations of the consequences of $g_N$\ on the
Higgs sector$^8$ and on the charginos and
neutralinos appear$^4$.
In this model the $\mu$\ parameter is replaced by $\mu=h_N{{v_N}\over
{\sqrt 2}}$ and we get in the chargino sector the same mass eigenstates,
diagonalizing matrices and angles as given in ref.6. Here the analytic
 results are well known. In
the  neutralino sector we have to perform numerical calculations. Only a few
limits give analytic expressions (e.g. gaugino masses equal to zero
or $\mu=0$).
In the following we will not consider these
special cases. We take the more realistic cases given in ref.8. There it was
shown that the RGEs drive the values of $h_N$\ and $k$\ to their infrared fixed
points if we start at the GUT scale with values of order 1 or larger and end up
at the weak scale. The results given there are $h_N\approx 0.87$\ and $k\approx
0.6$. Therefore only one new parameter r is left. In this talk we
take the two cases of r=0.1 (that is $\mu=15.3$\ GeV) and r=1
($\mu=153$\ GeV) for the values of $h_N$\ and $k$\ given above.
\hfill\break\indent In the following we consider three special interesting
cases. For all cases we will take the
well known  relation for the gaugino masses $m_{g_1}/m_{g_2}=
3\alpha_2/5\alpha_1$.
 We take into account the mass bound of the lightest higgsino-like
neutralino to be heavier than about 25 GeV$^{13}$. This bound is given by the
measurement of the invisible width of the $Z^0$-boson and from the ALEPH and
OPAL collaboration experiments$^{14}$, based on visible $Z^0$\ decays into
neutralinos.\hfill\break\indent
For the discussion of the mass matrices of the charginos and neutralinos
in the MSSM we refer the reader to the literature$^{6}$.
\hfill\break\indent
For the discussion of the mass matrices of the charginos and neutralinos
in the MSSM  without
the $\mu$\ parameter, but with an additional Higgs singlet N
we refer the reader to the first reference in ref.4.
There we have considered three different cases.
 In the first case we have taken r=0.1 (that is $\mu=15.3$\ GeV)
 and $\tan\beta=3/2$.
 In the other two cases we have r=1 with $\tan\beta= 3/2$\ and 4.
\hfill\break\indent In the next
section we want to present the result of the influences of the charginos and
neutralinos on the AMMM in the three cases considered above.
\hfill\break\vskip.2cm\noindent
{\bf III. THE ANOMALOUS
MAGNETIC MOMENT OF THE MUON IN SUSY MODELS EXTENDED BY AN ADDITIONAL HIGGS
SINGLET}\vskip.2cm\noindent In this section we only present our results, as
detailed calculations have been given elsewhere$^{4,5}$\
 and will not be repeated
here.  In the considered cases
we use for the mass of the  SU(2) doublet scalar muon
neutrino the lowest experimental bound of $m_{\tilde\nu_{\mu}}=30$\ GeV. From
the D-term in the  scalarpotential (note that our singlet here has no U(1)
charge and so does not contribute to the D-term) it follows that the SU(2)
doublet scalar muon mass is then given by$^9$
$m^2_{\tilde\mu}=m^2_{\tilde\nu_{\mu}}+m^2_W
{{v^2_2-v^2_1}\over{v^2_1+v^2_2}}$,
which gives us  $m_{\tilde\mu}\approx 59$\ GeV for $\tan\beta=3/2$\ and 82 GeV
for  $\tan\beta=4$.
We also assume $m_{\tilde\mu_L}=m_{\tilde\mu_R}$ and we
neglect large mixing between the scalar partners of the left and right handed
muons
($m_{\mu}\ll m_{\tilde\mu}$).
\hfill\break\indent
In Fig.1 we have plotted the first case with $r=0.1$\
($\mu= 15.3$\ GeV) for $\tan\beta=3/2$. Here we see
that we get higher contributions to the AMMM if the gaugino masses get larger,
but we
have to remark that $m_{g_2}$\ has to be lower than about 100 GeV if we want to
have the
lightest chargino larger than 40 GeV. In this allowed gaugino mass region the
value for $\Delta a_{\mu}$\ goes from $-2.36\cdot 10^{-9}$\ up to $-2.14\cdot
10^{-9}$. In Fig.2 we have plotted the last two cases with $r=1$\
($\mu=153$\ GeV) for $\tan\beta=3/2$\ (solid line) and $\tan\beta=4$\ (dotted
line). For $\tan\beta=3/2$, $\Delta a_{\mu}$\ lies in the range of $-2.12\cdot
10^{-9}$\ to $-6.87\cdot 10^{-10}$\ and for $\tan\beta=4$\ in the range  of
$-1.42\cdot 10^{-9}$\ to $-5.28\cdot 10^{-10}$.
 As mentioned above we have to
take  $m_{g_2}\geq 100$\ GeV in both cases to have a lightest chargino with
mass
larger than 40 GeV. We stopped at 200 GeV for the gaugino mass of SU(2) since
for larger masses  $\Delta a_{\mu}$\ gets too small. Because the
uncertainty of the hadronic contribution to the AMMM is about $1.9\cdot
10^{-9}\ ^{15}$ a too large gaugino mass will make the SUSY contribution
undistinguishable from the uncertainties of the hadronic contribution.
 As a result, we see that an additional Higgs singlet N will
suppress the contribution from the MSSM if it
gets a large vev leading to a large $\mu$\ parameter in the
MSSM. This would also be true in the MSSM, if we had chosen $\mu=+153$\ GeV
instead
of $\mu=-153$\ GeV. The difference between the singlet model and the MSSM with
a large
$\mu$\ parameter lies basically in the neutralino sector, but the influence on
the AMMM
turns out to be the same. For a detailed discussion of the influence on the
AMMM within
the MSSM we refer the reader to ref.5.
%\vfill\break
%\phantom{a}
\vskip4.5cm
$$\vbox{\settabs2\columns
\+ \item{Fig.1}The contribution to the anomalous&\quad
\item{Fig.2}The contribution to the anomalous \cr
\+ magnetic moment of
 the muon $\Delta a_{\mu}$ as a &\quad magnetic moment of
 the muon $\Delta a_{\mu}$ as a\cr
\+ function of the gaugino mass of SU(2) in&\quad
function of the gaugino mass of SU(2) in\cr
\+ the case
r=0.1 ($\mu=-15.3$\ GeV) with &\quad the case
r=1 ($\mu=153$\ GeV) with\cr
\+ $\tan\beta=3/2$. We have taken $m_{\tilde\nu_{\mu}}=30$ &\quad
$\tan\beta=3/2$\ (solid line) and $\tan\beta=4$\cr
\+ GeV with $m_{\tilde\mu} =59$\ GeV.&\quad and (dotted line). For the
masses of the\cr
\+ &\quad scalar muon neutrino and scalar muon we\cr
\+ &\quad have taken $m_{\tilde\nu_{\mu}}=30$\ GeV with $m_{\tilde\mu}=59$\cr
\+ &\quad GeV for $\tan\beta=3/2$ and $m_{\tilde\mu}=82$\ GeV \cr
\+ &\quad for $\tan\beta=4$.\cr}$$
%\hfill\break\vskip.2cm\noindent
{\bf IV. CONCLUSIONS}\vskip.2cm In this
talk we have shown how an additional Higgs singlet N in the
supersymmetric standard model without a $\mu$\ parameter in the superpotential
affects the mass matrices of the charginos and neutralinos. We have also
shown the consequences for the anomalous magnetic moment of the muon if the
Higgs singlet involves a large vacuum expectation value. As  a result, we have
seen that a too large value $v_N$\ (that is a large $\mu$\ parameter) leads to
values of the contribution $\Delta a_{\mu}$\ to the anomalous magnetic moment
of
the muon which makes it impossible to distinguish it from the uncertainties of
the hadronic contributions. This is also true for the MSSM with a large
positive
$\mu$\ parameter. We conclude that if there will be an
experiment which gives a value for  $\Delta a_{\mu}^{\rm exp}$\ only a factor
10-20 better than the old value, the minimal  supersymmetric standard model
with
a small $\mu$\ parameter and not too large masses for the scalar muon
neutrino and scalar muon will be ruled out and we will get
strong lower limits for the gaugino masses if we have a large positive value
of the $\mu$\ parameter in the MSSM
or if we have a large vacuum expectation value $v_N$\ in a
supersymmetric model extended by a Higgs singlet N.\hfill\break\indent
As a final point we want to mention that we have also tried to include R-parity
breaking through a vacuum expectation value $v_{\tau}$\ of the scalar tau
neutrino within the extended supersymmetric model
considered here. But, due to the new variation of the Higgs potential with
respect to $v_{\tau}$, we get new constraints which cannot be fulfilled
$^{4,16}$.\hfill\break
\vskip.2cm
{\bf REFERENCES}\vskip.2cm
\item{[\ 1]}
J. Bailey et al., {\it Nucl.Phys.} {\bf B150}(1979),1.
\item{\phantom{[\ 1]}}
F.H. Compley, {\it Rep.Progr.Phys.} {\bf 42}(1979),1889
\item{[\ 2]}V.M. Hughes, {\it AIP Conference Proceedings,
Intersections between Part. and Nucl.Phys.}, Lake Louise, Canada 1986.
\item{[\ 3]}
B. Adeva et al, {\it Phys.Lett.} {\bf 233B}(1989),530.
\item{\phantom{[\ 7]}}D. Decamp et al, {\it Phys.Lett.}
{\bf 236B}(1990),86.
\item{\phantom{[\ 7]}}M. Akrawy et al, {\it Phys.Lett.}
{\bf 240B}(1990),261.
\item{\phantom{[\ 7]}}T. Barklov et al, SLAC PUB-5196(1990).
\item{[\ 4]}H.K\"onig, {\it Z. Phys.} {\bf C52}(1992),159
 ,{\it Mod. Phys. Lett.} {\bf A4}(1992),279.
\item{[\ 5]}I. Vendramin, {\it Nuovo Cimento} {\bf 101A}(1989),721.
\item{[\ 6]}
H.E. Haber and G.L. Kane, {\it Phys.Rep.} {\bf 117}(1985),75.
\item{\phantom{[\ 6]}}J.F. Gunion and H.E. Haber,
{\it Nucl.Phys.} {\bf B272}(1986),1.
\item{[\ 7]}H.P. Nilles, {\it Phys.Rep.} {\bf 110}(1984),1
\item{[\ 8]}M. Drees, {\it Int.J.Mod.Phys.}
{\bf  A4}(1989),3635.
\item{\phantom{[\ 8]}}J. Ellis et al, {\it Phys.Rev.}
{\bf D39}(1989),844.
\item{[\ 9]}L.E. Iba\~nez and C. L\'opez,
{\it Nucl.Phys.} {\bf B233}(1984),511,
{\it Nucl.Phys.} {\bf B218}(1983),514.
\item{[10]}
R.D. Peccei and H.R. Quinn,
{\it Phys.Rev.Lett.} {\bf 38}(1977),1444,
{\it Phys.Rev.} {\bf D16}(1977),
1791,Proceedings of the 1981 Kyoto Summer Institute.
\item{[11]}E. Witten, {\it Phys.Lett.} {\bf 155B}(1985),151,
{\it Nucl.Phys.} {\bf B258}(1985),75.
\item{ \phantom{[11]}}P. Candelas et al, {\it Nucl.Phys.}
{\bf B258}(1985),46.
\item{[12]}J.F. Gunion et al, {\it The Higgs Hunter's
Guide}, (Addison-Wesley, Redwood City, CA, 1990).
\item{[13]}L. Roszkowski,{\it Phys.Lett.} {\bf 252B}(1990),471.
\item{[14]}
J.F. Grivaz, talks presented at the 25th Rencontres de Moriond on
Electroweak Interactions and Unified Theories (March 1990), and at
PASCOS 90, North-Eastern University Boston, MA (March 1990).
\item{\phantom{[11]}}D. Decamp et al, {\it Phys.Lett.}
{\bf 244B}(1990),541.
\item{\phantom{[11]}}M.Z. Akrawy et al, {\it Phys.Lett.}
{\bf B248}(1990),211.
\item{[15]}T. Kinoshita et al, {\it Phys.Rev.} {\bf D41}(1990),593,
{\it Phys.Rev.}{\bf D31}(1985),2108.
\item{[16]}H. K\"onig, Ph.D.-thesis (German), unpublished.
\vfill\break
\end